\documentclass[sigconf]{acmart}
\AtBeginDocument{%
  }

\acmConference[Conference acronym 'XX]{Make sure to enter the correct
  conference title from your rights confirmation email}{June 03--05,
  2018}{Woodstock, NY}
\acmISBN{978-1-4503-XXXX-X/2018/06}

\copyrightyear{2025} \acmYear{2025} \setcopyright{cc} \setcctype{by}\setcctype{by} \acmConference[PoIDS '25]{The 1st ACM SIGSPATIAL International Workshop on Polar Data Science}{November 3--6, 2025}{Minneapolis, MN, USA} \acmBooktitle{The 1st ACM SIGSPATIAL International Workshop on Polar Data Science (PoIDS '25), November 3--6, 2025, Minneapolis, MN, USA} \acmDOI{10.1145/3764922.3771172} \acmISBN{979-8-4007-2185-4/2025/11}

\begin{document}

\title{Causal Feedback Discovery using Convergence Cross Mapping on Sea Ice Data}

\author{Francis Nji}
\affiliation{
  \institution{Department of Information Systems\\University of Maryland, Baltimore County}
  \city{Baltimore}
  \state{Maryland}
  \country{USA}
}
\author{Seraj Al Mahmud Mostafa}
\affiliation{
  \institution{Department of Information Systems\\University of Maryland, Baltimore County}
  \city{Baltimore}
  \state{Maryland}
  \country{USA}
}

\author{Jianwu Wang}
\affiliation{
  \institution{Department of Information Systems\\University of Maryland, Baltimore County}
  \city{Baltimore}
  \state{Maryland}
  \country{USA}
}

\renewcommand{\shortauthors}{Nji et al.}

\begin{abstract}
Identifying causal relationships in climate systems remains challenging due to nonlinear, coupled dynamics that limit the effectiveness of linear and stochastic causal discovery approaches. This study benchmarks Convergence Cross Mapping (CCM) against Granger causality, PCMCI, and VarLiNGAM using both synthetic datasets with ground truth causal links and 41 years of Arctic climate data (1979--2021). Unlike stochastic models that rely on autoregressive residual dependence, CCM leverages Takens' state-space reconstruction and delay-embedding to reconstruct attractor manifolds from time series. Cross mapping between reconstructed manifolds exploits deterministic signatures of causation, enabling the detection of weak and bidirectional causal links that linear models fail to resolve. Results demonstrate that CCM achieves higher specificity and fewer false positives on synthetic benchmarks, while maintaining robustness under observational noise and limited sample lengths. On Arctic data, CCM reveals significant causal interactions between sea ice extent and atmospheric variables like specific humidity, longwave radiation, and surface temperature with a $p$-value of $0.009$, supporting ice-albedo feedbacks and moisture-radiation couplings central to Arctic amplification. In contrast, stochastic approaches miss these nonlinear dependencies or infer spurious causal relations. This work establishes CCM as a robust causal inference tool for nonlinear climate dynamics and provides the first systematic benchmarking framework for method selection in climate research.\end{abstract}

\ccsdesc[500]{Applied computing~Earth and atmospheric sciences}
\keywords{Causality, Causal Feedback, Convergence Cross Mapping, CCM, Granger Causality, PCMCI, VarLiNGAM.}
\begin{teaserfigure}
  \centering
  \color{red}
Note: The data in Table~1 in the published ACM version (https://dl.acm.org/doi/pdf/10.1145/3764922.3771172) was incorrect; it has been corrected in this version.
\end{teaserfigure}

\received{10 August 2025}
\received[revised]{25 September 2025}
\received[accepted]{6 October 2025}

\maketitle

\section{Introduction}
The areal extent, concentration, and thickness of sea ice in the Arctic ocean and adjacent seas have dramatically decreased over recent decades \cite{stroeve2012arctic}. This decline has significantly increased heat flux from the ocean to the atmosphere in autumn and early winter, leading to locally increased air temperatures, moisture, and cloud cover, while reducing static stability in the lower troposphere \cite{stroeve2012arctic}. Studies suggest that with advancing global warming, cold winters in mid-latitude continents may become uncommon in the second half of the twenty-first century \cite{blackport2019minimal}. Additionally, recent findings have linked sea ice decline to summer precipitation patterns in Europe, the Mediterranean, and East Asia \cite{vihma2014effects, ali2022benchmarking}. The reduction of Arctic ice sheets is a prominent indicator of warming, evidenced by a nearly 50\% decrease in ice extent since 1979 \cite{stammerjohn2012regions}. As the Arctic sea ice cover diminishes, the annual mean 2-m air temperature has increased across nearly all weather stations north of 60N, particularly in coastal and archipelago areas surrounding the Arctic Ocean \cite{polyakov2012recent}.

The effects of sea ice decline can be categorized into local and remote effects depending on the season. Local effects occur in regions that have lost sea ice cover while remote effects impact areas that have historically been ice-free. The expected direct local effects include increased turbulent fluxes of sensible and latent heat from the ocean to the atmosphere, as well as enhanced longwave radiation emitted by the sea surface. Understanding the remote effects is more complex, as it involves discerning how mid-latitude circulation responds to Arctic changes in the absence of other influencing factors. 
Interactions among atmospheric time series variables are inherently complex and nonlinear due to the intricate and often unpredictable dependencies that exist between them. In the atmosphere, factors such as temperature, humidity, sea ice extent, and radiation do not evolve in isolation; rather, they interact through multifaceted pathways. For example, an increase in temperature can accelerate sea ice melt, which reduces surface albedo and subsequently amplifies local and regional warming, thereby altering weather patterns. The abundance of feedback loops and deterministic components in atmospheric processes causes information about these variables to become inseparably intertwined. Identifying causal relationships is therefore essential to disentangle these interactions, improve predictive understanding, and design effective interventions.
These interactions are fundamentally nonlinear because small perturbations in one variable can trigger disproportionately large responses in others, rendering the system highly sensitive to initial conditions. Moreover, feedback mechanisms where the output of a process feeds back into its own input further amplify complexity and nonlinearity. In this context, uncovering causal relationships and feedback loops is critical for advancing climate science, enabling more accurate forecasts, and informing targeted mitigation and adaptation strategies. 
Recent work applying Convergent Cross Mapping (CCM) to Arctic datasets has demonstrated its ability to uncover nonlinear feedbacks between sea ice, atmospheric moisture, and radiation processes, validating its use in polar climate research \cite{su2023uncovering, gao2023causal}. While controlled randomized experiments are ideal for efficient causal discovery, they are often unethical, costly, or technically unfeasible \cite{hoyer2008nonlinear}. As a result, research has focused on causal discovery methods that infer relationships from uncontrolled data (i.e observations collected from natural or real-world processes with many unknown and unmeasured influencing factors). Although stochastic causal models such as Granger causality, PCMCI and VarLiNGAM (they explicitly model causal relationships as probabilistic dependencies driven by random processes) are frequently employed due to their simplicity, they often fail to accurately capture complex, non-linear causal relationships because they are fundamentally constrained by their assumptions of linearity, stationarity, and acyclicity whereas atmospheric and climate dynamics are nonlinear: feedbacks (where processes amplify or dampen their own effects), thresholds (points beyond which behavior qualitatively shifts), or bifurcations (where the system’s regime structure changes abruptly) \cite{rial2004nonlinearities}, nonstationary (climate change alters statistical properties over time), cyclic and feedback-driven (e.g., ice–albedo, ocean–atmosphere coupling) and finally high-dimensional with hidden confounders. In nonlinear systems every coupled variable carry information about others. This implies that variables cannot be fully removed from the system for analysis, and this further violates the linearity assumption made by stochastic causal models.

In this study, we apply Convergent Cross Mapping (CCM) to complex multivariate spatiotemporal Arctic climate data to estimate dynamic coupling by detecting nonlinear causal feedback loops and weak interactions often missed by traditional stochastic causal models. Our results demonstrate that CCM effectively identifies and quantifies causal relationships in the Arctic climate system with non-separable variables, including weak to moderate interactions that stochastic approaches frequently overlook. Moreover, CCM operates independently of predictive models, enabling more robust and model-free causal inference.

\section{Background and Related Work}

\subsection{Background} 
Detecting causal drivers and regime shifts in Arctic climate systems presents a critical challenge due to nonlinear feedbacks, high-dimensional couplings, and non-separable spatiotemporal dynamics. Sea ice extent, atmospheric circulation, and oceanic heat fluxes interact through bidirectional feedbacks, complicating conventional stochastic causal discovery methods. Convergent Cross Mapping (CCM) addresses these limitations by reconstructing state-space manifolds from observed time series and assessing whether the historical states of one variable can reliably estimate another, enabling detection of weak and asymmetric causal links without relying on predictive models. Concurrently, advances in deep unsupervised spatiotemporal representation learning \cite{10825871, nji2024evaluation, faruque2023deep} demonstrate the ability to uncover latent temporal regimes, structural shifts, and evolving patterns in high-dimensional climate data. While representation learning identifies emergent regime structures, CCM provides mechanistic insight by quantifying the directional couplings that drive these patterns, revealing feedback loops and causal pathways underlying Arctic amplification. Natural systems, such as atmospheric processes, are inherently complex and nonlinear. Linear correlation-based approaches are often inadequate for these systems because correlations can arise without true causal relationships, while genuine causal influences may exist even in the absence of measurable correlation. For causality detection, if $\sigma^2(X | U) < \sigma^2(X | U - \overline{Y})$, we say that $Y$ is causing $X$, denoted by $Y_t \Rightarrow X_t$. This means $Y_t$ causes $X_t$ if we can predict $X_t$ better using all available information than without $Y_t$.

CCM leverages \textit{Takens’ Theorem}~\cite{takens2006detecting}, which states that the hidden dynamics of a complex system can be reconstructed from a single observed time series using time-delayed copies of that variable. This process, known as \textit{State-Space Reconstruction}, unfolds the system’s trajectory into a multidimensional space where each dimension represents the system’s state at different time lags. Within this reconstructed space, the system evolves along an \textit{Attractor Manifold}---a geometric surface that captures the stable, recurring patterns toward which the system naturally tends over time. CCM builds upon this principle by constructing ``shadow manifolds’’ from observed variables that preserve the same topological structure as the original system. The embedding dimension \(E\) determines how many delayed coordinates are required to fully capture the system’s degrees of freedom, while the time delay \(\tau\) controls how separated these snapshots are in time. Choosing appropriate \(E\) and \(\tau\) ensures that the manifold is unfolded correctly—too small and it collapses, too large and it becomes noisy. Once reconstructed, cross mapping exploits the fact that if variable \(X\) causally influences \(Y\), then traces of \(X\)’s dynamics are embedded in \(Y\)’s manifold, allowing \(X\) to be predicted from \(Y\). Unlike traditional methods such as Granger causality, which focus on predictive improvement, CCM identifies direct state-space dependence, making it robust for nonlinear and noisy systems. Applications in climate science confirm that CCM successfully identifies causal feedbacks between sea-ice extent, temperature, and radiative processes~\cite{su2023uncovering, gao2023causal}, offering a powerful framework for disentangling drivers of Arctic sea-ice decline and associated regime shifts.

\subsection{Related Work}
Convergent Cross Mapping is designed to identify and quantify causalities in systems whose variables are not separable. Clark et al. proposed Multispatial CCM which combine the existing techniques of CCM and dewdrop regression to build a novel test of causal relations that leverages spatial replication to detect causal relationships between processes \cite{clark2015spatial}. While the ability of their proposed model was tested on simulated and real-world ecological data, their performance on real world spatiotemporal data is yet unexplored. Brouwer et al. \cite{de2020latent}, proposed Latent CCM which uses reconstruction between latent processes of dynamical systems to infer causality between short and sporadic time series. Although their proposed seemed to work well on neural activity data, its application to climate data has not yet been explored.  Ye et. al. \cite{ye2015distinguishing}, demonstrate the ability of CCM to identify different time-delayed interactions, distinguish between synchrony induced by strong unidirectional-forcing and true bidirectional causality, and resolve transitive causal chains. Their application is limited to model simulations, a laboratory predator-prey experiment, temperature and greenhouse gas reconstructions from the Vostok ice core, and longterm ecological time series collected in the Southern California Bight with no known application to complex multidimensional atmospheric data.  Wang et. al. applied CCM to explore the causality between soil moisture and precipitation over low- and mid- latitude regions \cite{wang2018detecting}. Similarly, recent Arctic-focused studies demonstrate CCM’s strength in resolving feedback loops between sea ice extent and atmospheric drivers such as humidity and radiation \cite{gao2023causal, ye2015distinguishing}.
\section{Methodology}

\subsection{Synthetic Data Generation}
\begin{equation} \label{formula}
\resizebox{\columnwidth}{!}{$
\begin{aligned}
\text{Variable 1: } & S_1(t) = 0.125\sqrt{2}\exp\left(-\frac{S_1(t-1)^2}{2}\right) + 0.3\sqrt{2}\exp\left(-\frac{S_1(t)^2}{2}\right) \\
& \quad + 0.2\exp\left(-\frac{S_5(t-3)^2}{2}\right) + 0.2\exp\left(-\frac{S_6(t)^2}{2}\right) + \varepsilon_1 \\
\text{Variable 2: } & S_2(t) = 1.2\exp\left(-\frac{S_1(t-1)^2}{2}\right) + 0.2\exp\left(-\frac{S_1(t-2)^2}{2}\right) \\
& \quad + 0.2\exp\left(-\frac{S_5(t-2)^2}{2}\right) + 0.2\exp\left(-\frac{S_3(t-1)^2}{2}\right) + \varepsilon_2 \\
\text{Variable 3: } & S_3(t) = -1.05\exp\left(-\frac{S_1(t-1)^2}{2}\right) + 0.2\exp\left(-\frac{S_3(t)^2}{2}\right) \\
& \quad + 0.2\exp\left(-\frac{S_2(t-2)^2}{2}\right) + 0.2\exp\left(-\frac{S_6(t-2)^2}{2}\right) + \varepsilon_3 \\
\text{Variable 4: } & S_4(t) = -1.15\exp\left(-\frac{S_1(t-1)^2}{2}\right) + 0.2\sqrt{2}\exp\left(-\frac{S_4(t-1)^2}{2}\right) \\
& \quad + 1.35\exp\left(-\frac{S_3(t-1)^2}{2}\right) + \varepsilon_4 \\
\text{Variable 5: } & S_5(t) = -1.15\exp\left(-\frac{S_1(t-3)^2}{2}\right) + 0.2\sqrt{2}\exp\left(-\frac{S_2(t-2)^2}{2}\right) \\
& \quad + 1.35\exp\left(-\frac{S_3(t-1)^2}{2}\right) + \varepsilon_5 \\
\text{Variable 6: } & S_6(t) = -1.05\exp\left(-\frac{S_1(t-1)^2}{2}\right) + 0.2\exp\left(-\frac{S_3(t)^2}{2}\right) \\
& \quad + 0.2\exp\left(-\frac{S_2(t-2)^2}{2}\right) + 0.2\exp\left(-\frac{S_7(t)^2}{2}\right) + \varepsilon_6 \\
\text{Variable 7: } & S_7(t) = -1.05\exp\left(-\frac{S_4(t-2)^2}{2}\right) + 0.2\exp\left(-\frac{S_7(t)^2}{2}\right) \\
& \quad + 0.2\exp\left(-\frac{S_5(t-3)^2}{2}\right) + 0.2\exp\left(-\frac{S_6(t)^2}{2}\right) + \varepsilon_7 \\
\text{Variable 8: } & S_8(t) = -1.05\exp\left(-\frac{S_7(t-2)^2}{2}\right) + 0.2\exp\left(-\frac{S_8(t)^2}{2}\right) \\
& \quad + 0.2\exp\left(-\frac{S_6(t-1)^2}{2}\right) + 0.2\exp\left(-\frac{S_2(t-3)^2}{2}\right) + \varepsilon_8,
\end{aligned}
$}
\end{equation}
Our goal is to detect both linear and non-linear causal relationships, including weak and bidirectional feedback loops, among time series atmospheric variables. To achieve this, we evaluate a set of causal discovery methods without bias: stochastic models (Granger causality \cite{barnett2009granger}, PCMCI \cite{runge2020discovering}, VarLiNGAM \cite{shimizu2006linear}) and the nonlinear Convergent Cross Mapping (CCM), applied to both synthetic and real-world Arctic climate data.
Since ground-truth causal relationships are typically unavailable in empirical settings, we constructed a synthetic dataset to rigorously assess the accuracy of CCM in both causal discovery and feedback loop identification. The dataset comprises eight time series variables, denoted as $S_1, \ldots, S_8$, each containing 100{,}000 observations and governed by a lag structure of up to order three. To enable systematic evaluation of feedback detection, we explicitly incorporated bidirectional causal dependencies into the generative process, thereby providing controlled conditions under which the model’s capacity to recover feedback mechanisms can be validated.

\begin{figure}[ht!]
   \centering
   \includegraphics[width=\linewidth]{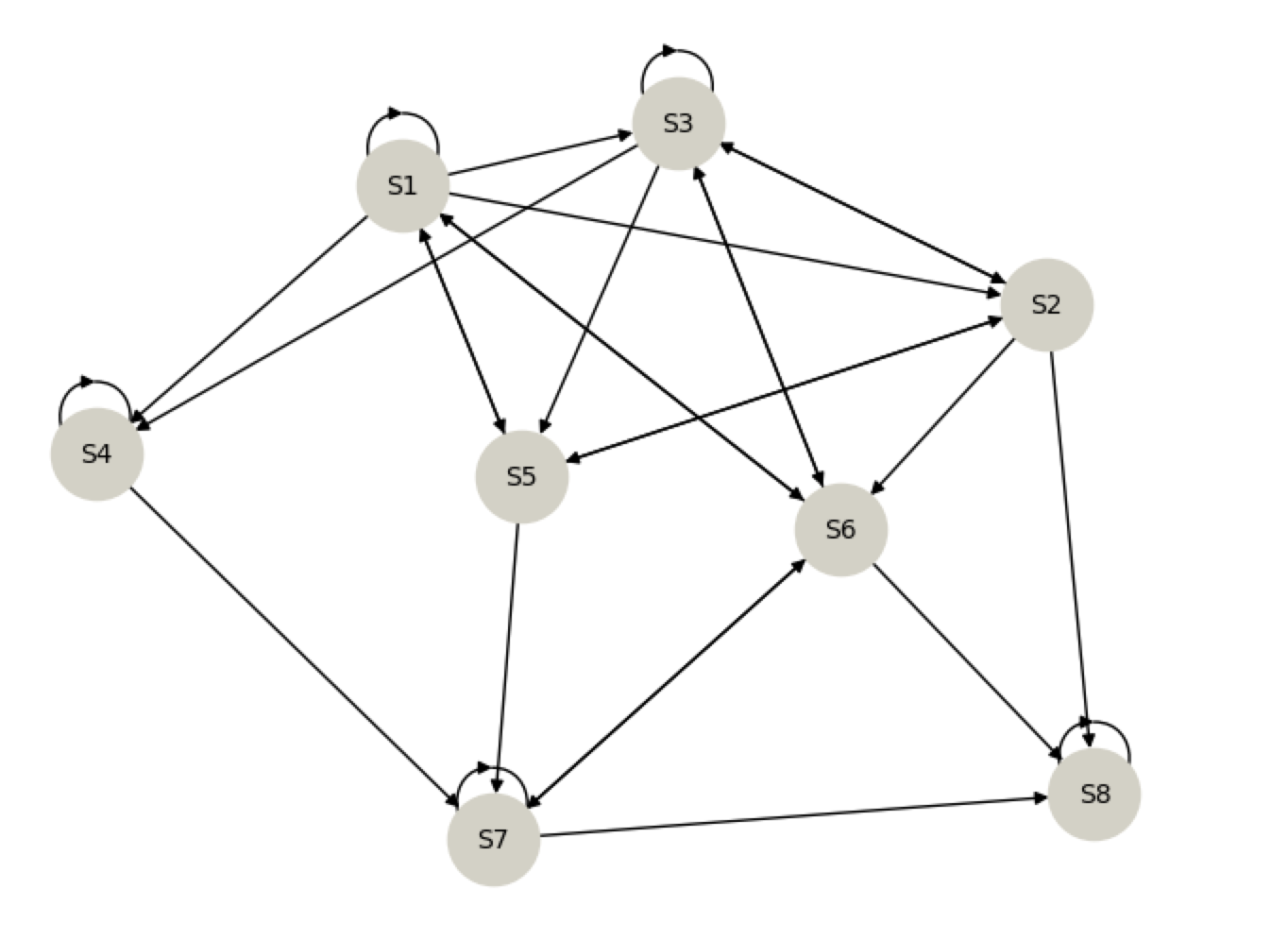}
   \caption{True causal graph of the synthetic dataset.}
   \Description{}
   \label{fig:true_cg}
\end{figure}

Figure~\ref{fig:true_cg} shows the corresponding causal graph, which includes forward edges and 6 introduced feedback edges:
$S_1 \Leftrightarrow S_5, S_1 \Leftrightarrow S_6, S_3 \Leftrightarrow S_6, S_3 \Leftrightarrow S_2, S_2 \Leftrightarrow S_5, S_6 \Leftrightarrow S_7.$ The graph is sparse with varying edge strengths.

\subsection{Real-World Dataset and Preprocessing}
\begin{table}[ht!]
\centering
\resizebox{0.5\textwidth}{!}{%
\begin{tabular}{|c| l| c| c|}
\hline
\textbf{No.} & \textbf{Variable} & \textbf{Range} & \textbf{Source} \\
\hline
1  & surface pressure & [40k, 110k] & ERA5 \\
2   & specific humidity & [0, 0.1] & ERA5 \\
3  & 2m temperature & [200, 350] & ERA5 \\
4 & shortwave radiation & [0, 1500] & ERA5 \\
5  & longwave radiation & [0, 700] & ERA5 \\
6  & rainfall & [0, 800] & ERA5 \\
7  & snowfall & [0, 200] & ERA5 \\
8  & sea surface temperature & [200, 350] & ERA5 \\
9  & 10m u-wind & [4800, 5700] & ERA5 \\
10 & sea ice extent & [3$\times$10$^{6}$, 14$\times$10$^{6}$] & NSIDC \\
\hline
\end{tabular}%
}
\caption{Arctic Sea Ice Data from 1979 – 2021.}
\label{tab:my-table3}
\end{table}

Table~\ref{tab:my-table3} consists of real-world Arctic Sea Ice Data from 1979 – 2021 with 15,160 temporal daily records and provides high-resolution spatiotemporal climate variables over the Arctic domain. The temporal resolution is hourly and aggregated to daily averages, allowing fine-grained analysis of day-to-day and seasonal variability. Spatially, the data are gridded on a 1° × 25 km horizontal resolution, representing a detailed depiction of mesoscale Arctic processes. Each grid cell records atmospheric and surface parameters - such as temperature, snow depth, wind components, and radiation fluxes capturing both temporal evolution and spatial heterogeneity across Greenland and the Arctic region. This high-resolution spatiotemporal coverage enables accurate regional climate and snowmelt pattern analyses.\\  
\textbf{Data preprocessing} included exploratory analysis, time series preprocessing, and normalization. We tested stationarity using the Augmented Dickey-Fuller~\cite{dickey1979distribution} and KPSS tests~\cite{kwiatkowski1992testing}. Seasonal and cyclical effects were removed via additive decomposition ($T_t + C_t + S_t + R_t$). Exponential smoothing reduced noise, and standardization ensured mean~0 and variance~1, using \texttt{StandardScaler()} from \texttt{scikit-learn}.

\subsection{Benchmark Time Series Causal Discovery Methods}

\noindent \textbf{Granger Causality.} A time series $X_t$ is said to \emph{Granger-cause} another time series $Y_t$ if the past values of $X_t$ contain statistically significant information that helps predict future values of $Y_t$, beyond the information already contained in the past values of $Y_t$ alone \cite{shojaie2022granger}.

\noindent \textbf{PCMCI (Peter--Clark Momentary Conditional Independence).} is a causal discovery algorithm for high-dimensional time series. It extends the PC algorithm to temporal data by combining conditional independence tests with momentary conditional independence (MCI) tests. PCMCI first reduces the candidate parent set using a condition-selection step, then applies rigorous independence testing to infer causal links \cite{runge2019inferring}.

\noindent \textbf{VarLiNGAM} is a Linear Non-Gaussian Acyclic Model for vector autoregressions (VAR), extending LiNGAM to time series. It assumes each variable is a linear function of contemporaneous causes and its own (and others’) past lags, with non-Gaussian, independent errors. Causal discovery separates contemporaneous effects (a DAG) from lagged effects, leveraging ICA-like independence and acyclicity constraints. Estimation fits a VAR, extracts residuals, then applies LiNGAM to residual contemporaneous relations to infer the causal ordering \cite{jiao2024optimizing}.
\subsection{Hardware Setup}
All models were executed on an AWS cloud environment with 20GB S3 storage and an \texttt{ml.g4dn.xlarge} GPU (30 GB). Additional experiments were verified locally on macOS Sonoma (v14.4.1, M1 Pro chip, 16 GB RAM). The same Python libraries were applied across all models to ensure consistency.

\noindent \textbf{Proposed CCM Approach.}
\begin{figure}[ht!]
   \centering
   \includegraphics[width=\linewidth]{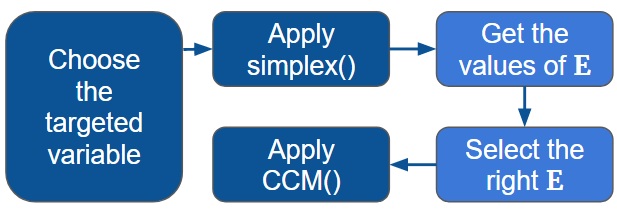}
   \caption{CCM workflow for feedback detection.}
   \Description{}
   \label{fig:ccm}
\end{figure}
We employed the \texttt{rEDM} package~\cite{van2015causal}, following the workflow in Figure~\ref{fig:ccm}. Key steps include:  
(i) selecting a target variable and computing embeddings $E$ using the \textit{simplex forecasting} function (even for a chaotic time series, future values can be predicted from the behaviour of similar past values)
(ii) choosing optimal $E$ based on forecast skill $\rho$,  
(iii) applying CCM to detect causal influence and feedback among pairs of variables.  
\section{Experiments and Results}

\subsection{Synthetic Data Experiments}
To provide a controlled benchmark with well-defined ground truth, we initially evaluated all four causal discovery methods on a synthetic dataset. This experimental setting ensures objective assessment of each approach’s accuracy and robustness in identifying both forward and feedback causal dependencies.  
\subsubsection{Granger Causality}
\begin{figure}[htp]
   \centering
   \includegraphics[width=\linewidth]{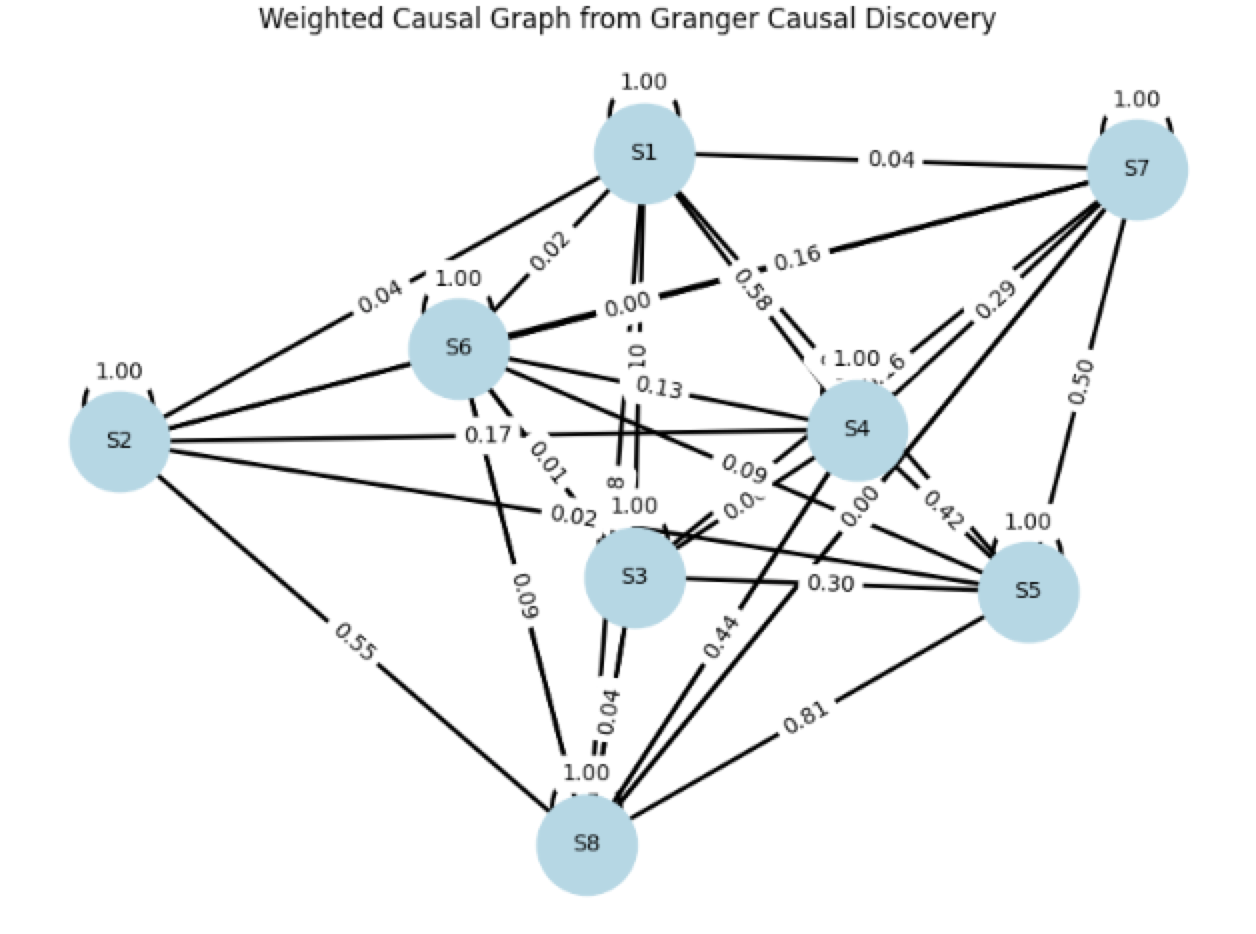}
   \caption{Learned causal graph from synthetic data using Granger causality method.}
   \Description{}
   \label{fig:gc_syn_graph}
\end{figure}

Granger causality is able to detect a subset of true causal links. Figure~\ref{fig:gc_syn_graph} illustrates identified edges in the synthetic system. While effective for strong linear lagged effects, the method failed to capture most bidirectional causal feedback loops.

\subsubsection{PCMCI} 
\begin{figure}[ht!]
   \centering
   \includegraphics[width=\linewidth]{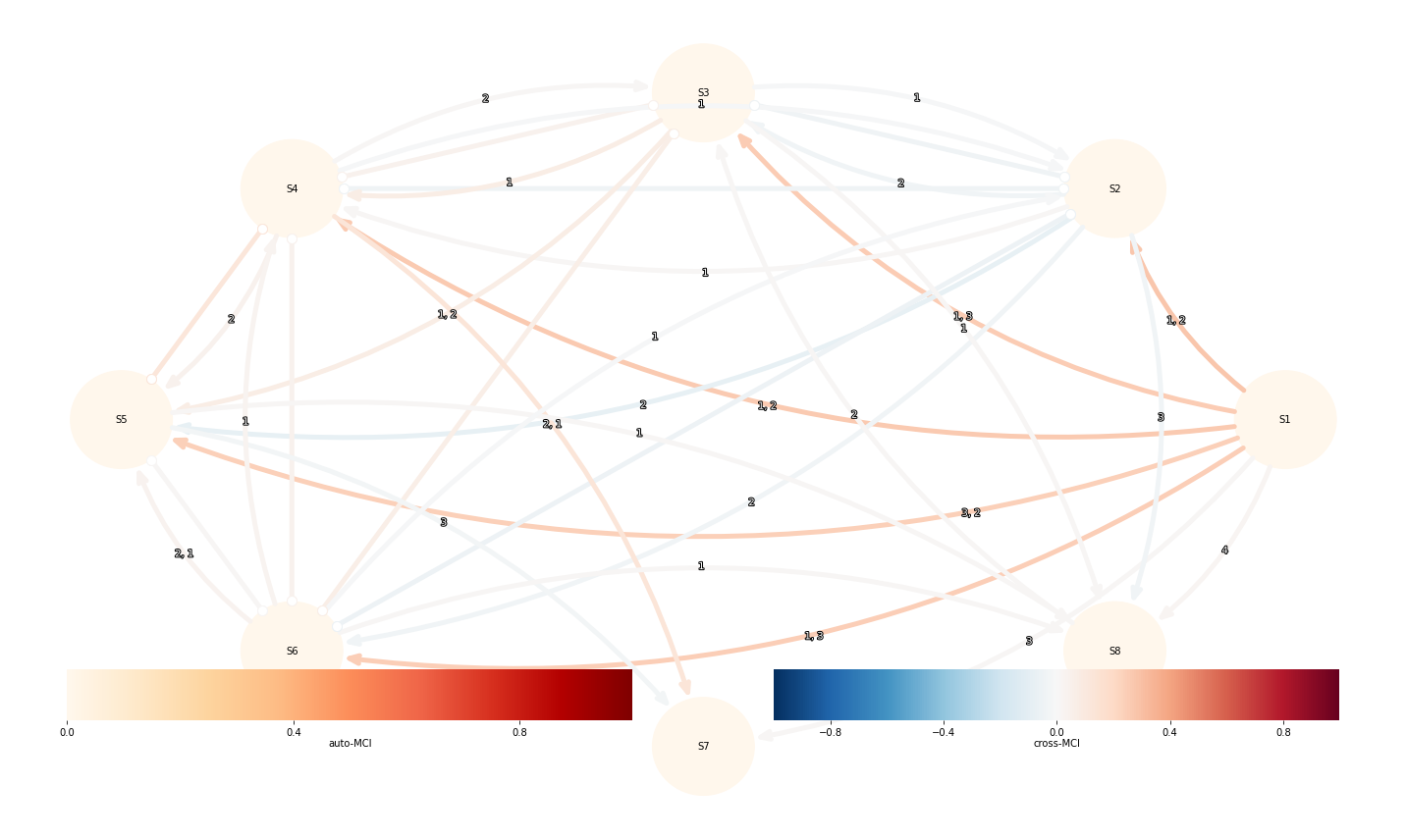}
   \caption{Learned causal graph from synthetic data using PCMCI.}
   \Description{}
   \label{fig:pcmci_m}
\end{figure}
As illustrated in Figure~\ref{fig:pcmci_m}, PCMCI demonstrated robustness in recovering several ground-truth causal dependencies, including 
$S_1 \rightarrow S_2, S_3, S_4, S_5, S_6$ and $S_3 \rightarrow S_5, S_6$. 
However, the method also introduced spurious associations and exhibited limited sensitivity to weaker feedback mechanisms. 
This suggests that while PCMCI is effective in detecting dominant causal drivers, its reliability diminishes when capturing subtler or reciprocal interactions within complex systems.

\subsubsection{VarLiNGAM} 
\begin{figure}[ht!]
   \centering
   \includegraphics[width=\linewidth]{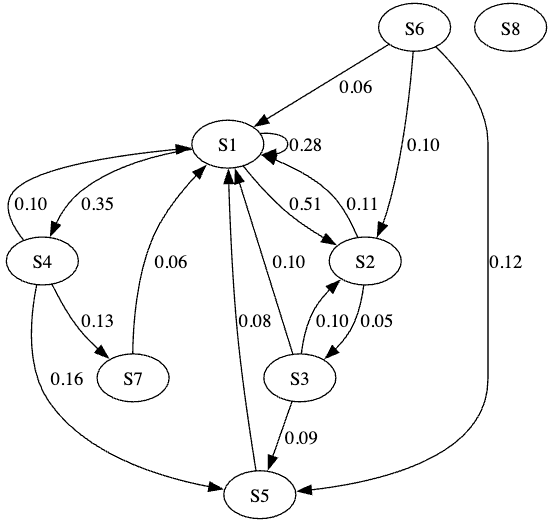}
   \caption{Learned causal graph from synthetic data using VarLiNGAM.}
   \Description{}
   \label{fig:VarLiNGAM}
\end{figure}

It integrates LiNGAM with vector autoregressive models, reported significant causal effects up to lag 2. Figure~\ref{fig:VarLiNGAM} is the learned causal graph from the synthetic data. Nodes are represented only with their variable names with lags collapsed for visualization purposes. We applied a p-value lower threshold of 0.05 and edges below this threshold were dropped. Overall, the model was consistent with the synthetic design but biased toward linear structures.

\subsubsection{Convergent Cross Mapping (CCM)}
\begin{figure}[htp]
   \centering
   \includegraphics[width=\linewidth]{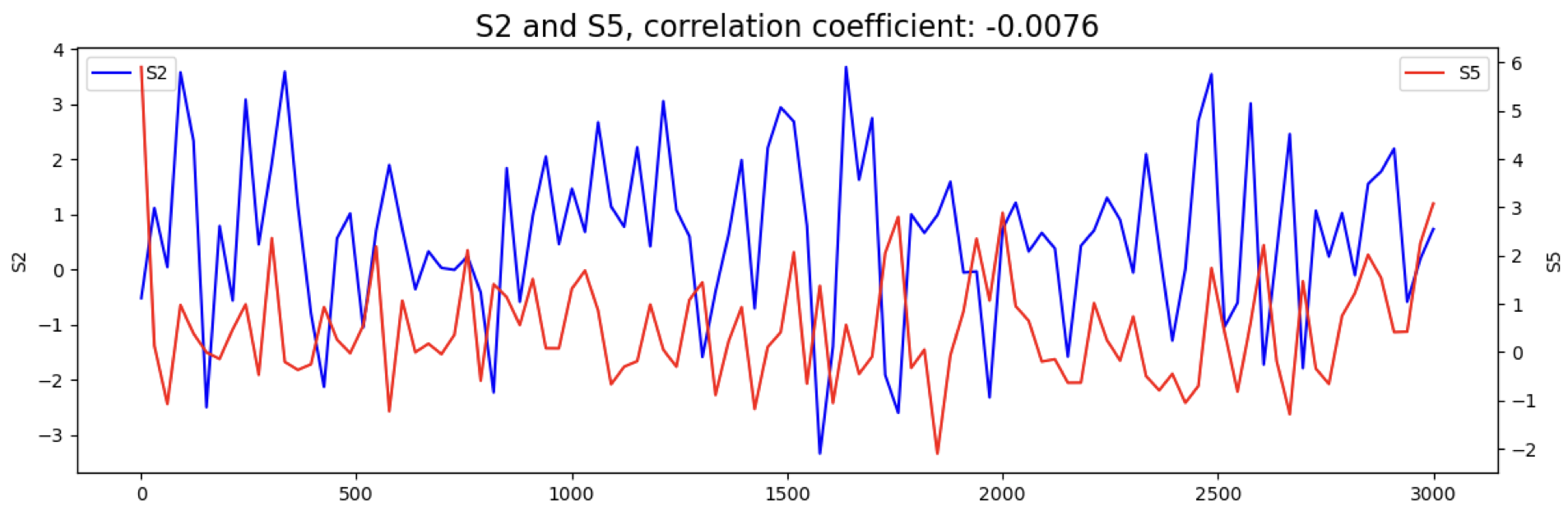}
   \caption{Learned feedback by CCM between $S_2$ and $S_5$.}
   \Description{}
   \label{fig:s5-s2}
\end{figure}

\begin{figure}[htp]
   \centering
   \includegraphics[width=\linewidth]{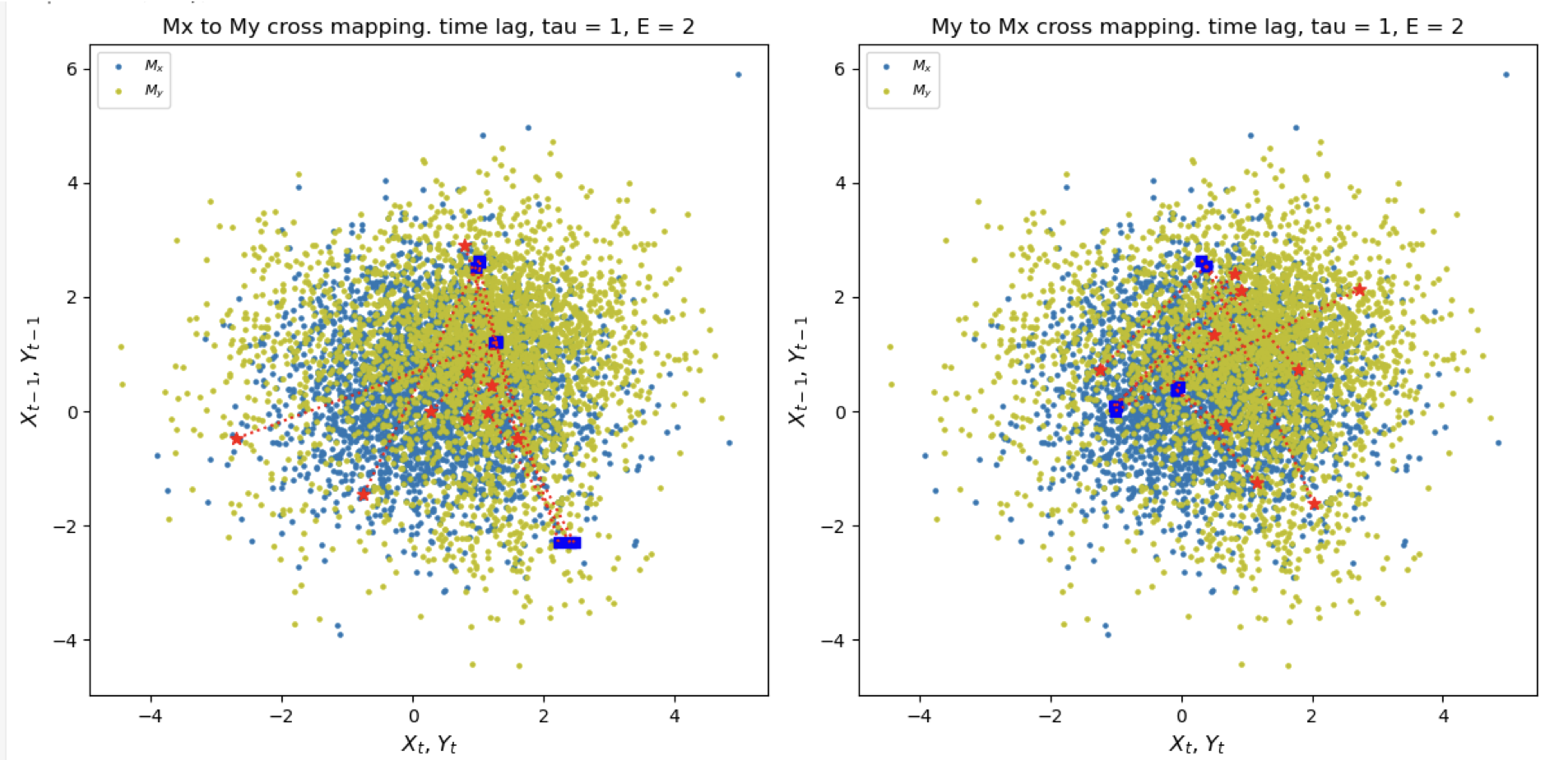}
   \caption{Shadow manifolds $M_x$ and $M_y$ for different $\tau$.}
   \Description{}
   \label{fig:cross-map}
\end{figure}

\begin{figure}[htp]
   \centering
   \includegraphics[width=\linewidth]{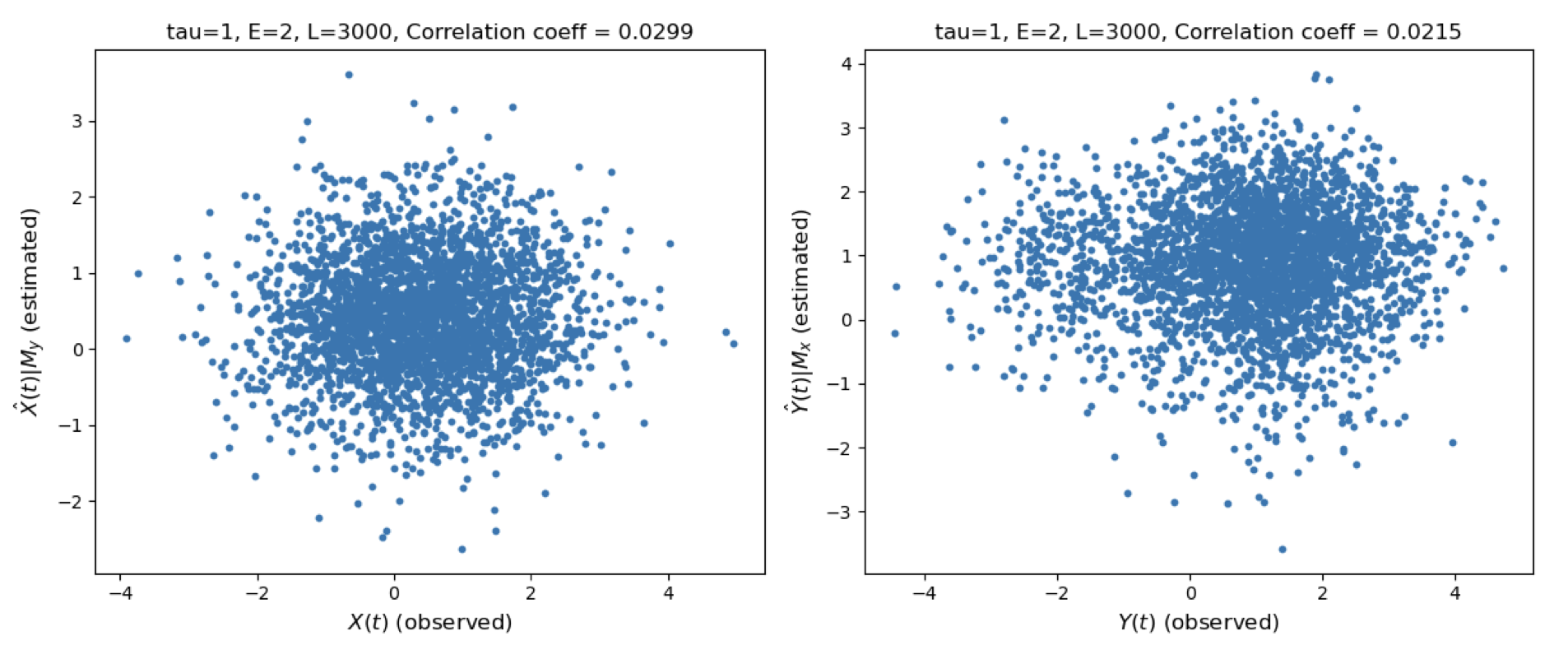}
   \caption{Cross mapping performance ($S_2$ vs. $S_5$).}
   \Description{}
   \label{fig:s2-s5-estimated}
\end{figure}

\begin{figure}[htp]
   \centering
   \includegraphics[width=\linewidth]{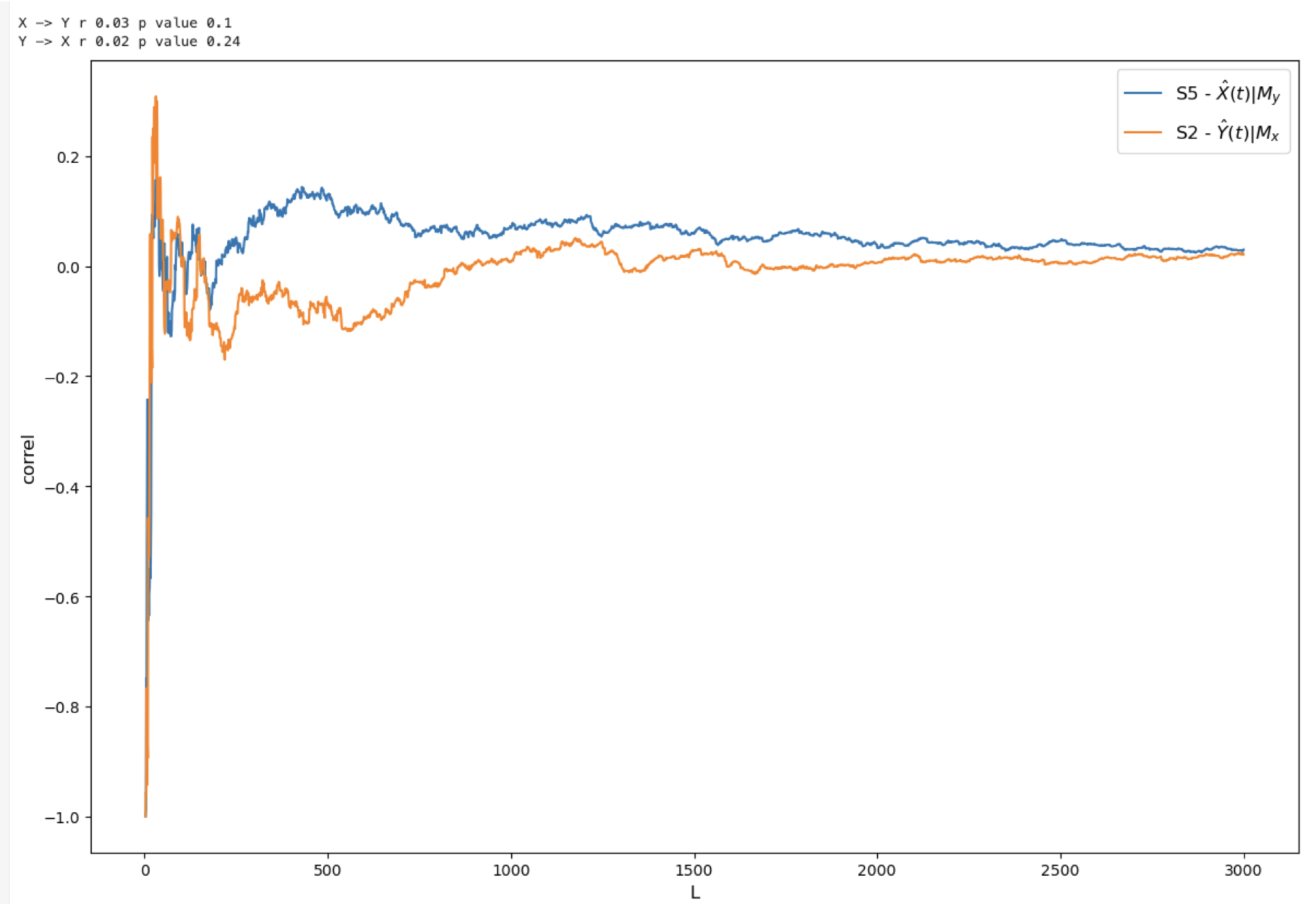}
   \caption{Convergence of CCM for $S_2 \Leftrightarrow S_5$.}
   \Description{}
   \label{fig:s2-s5}
\end{figure}

CCM consistently identified the designed bidirectional feedbacks. Figures~\ref{fig:s5-s2}--Figures~\ref{fig:s2-s5} illustrate results for $S_2 \leftrightarrow S_5$:  
shadow manifolds (Figure~\ref{fig:cross-map}), estimated cross-mapping (Figure~\ref{fig:s2-s5-estimated}), and convergence plots (Figure~\ref{fig:s2-s5}). Unlike Granger and PCMCI, CCM recovered weak causal links with high specificity.

\noindent
The shadow manifolds illustrated by Figure \ref{fig:cross-map} shows the cross mapping of points from one manifold (\textcolor{blue}{blue box}) to the corresponding points on another manifold (\textcolor{red}{red star}). If values from one manifold, for example \(M_X\), produce a widely \emph{spread-out} cross mapping to values on another manifold \(M_Y\), this indicates that we are unable to precisely predict \(M_Y\) (or the corresponding variable \(Y\)) given \(M_X\). In such a case, limited information from \(Y\) is embedded in \(X\), suggesting that while \(Y\) may drive \(X\), the causal influence is weak. Conversely, the \emph{stronger} the causality between variables, the \emph{narrower} the cross mapping becomes. For instance, cross mapping \(M_X\) from \(M_Y\) (as shown in the left panel) results in a relatively narrow mapping, implying that \(X\) has a stronger influence on \(Y\). While a narrow cross convergence pattern can be an indicator of potential causality, it represents only one of the two key criteria required to establish causal relationships in Convergent Cross Mapping (CCM). The second criterion—\textbf{convergence}—requires that the accuracy of cross mapping improves as the library size increases.

\noindent
We measure cross mapping using the \textbf{correlation coefficient} between the observed and predicted values. A stronger correlation indicates a stronger causal relationship. In Figure \ref{fig:s2-s5-estimated} the left panel illustrates the performance of cross mapping from manifold \(M_Y\) to manifold \(M_X\) in predicting \(X(t)\). There exist little correlation. This means that we cannot accurately estimate \(X\) given observations of \(Y\), implying that \(Y\) contains insufficient information about the dynamics of \(X\). In such a case, we infer that \(X\) \emph{weakly influences} or \emph{causes} \(Y\).


\begin{figure}[htp]
   \centering
   \includegraphics[width=\linewidth]{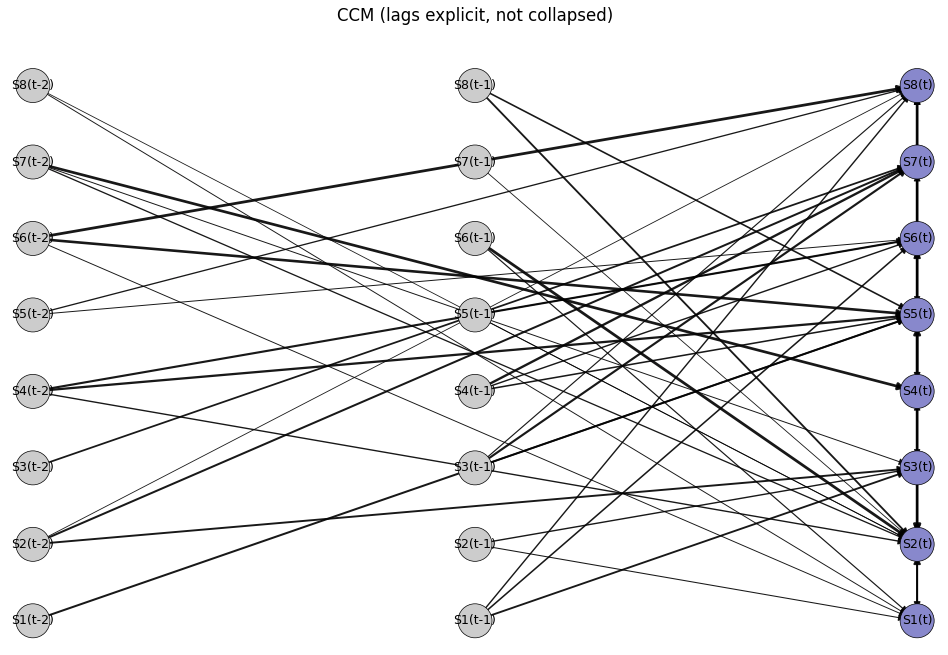}
   \caption{Learned causal graph with CCM on synthetic data}
   \Description{}
   \label{fig:ccm_syn}
\end{figure}
Convergence means we are able to improve cross mapping accuracy the longer the period \(L\) we consider. This happens only when we are able to enhance our reconstruction of the shared attractor between two variables using more data. In Figure \ref{fig:s2-s5}, we find that both cross mappings converge, however one converges more than the other. We have satisfied the two criteria for CCM: convergence and cross mapping. We can thus say weak causality exists. However, the magnitudes of causality going from \(S5\) or \(S2\) are different.

\subsection{Real-World Arctic Climate Experiments}

Next, we applied the abovementioned time series causal discovery methods to our Arctic climate dataset as shown in Table~\ref{tab:my-table3}.

\subsubsection{Granger Causality}
\begin{figure}[ht!]
   \centering
   \includegraphics[width=\linewidth]{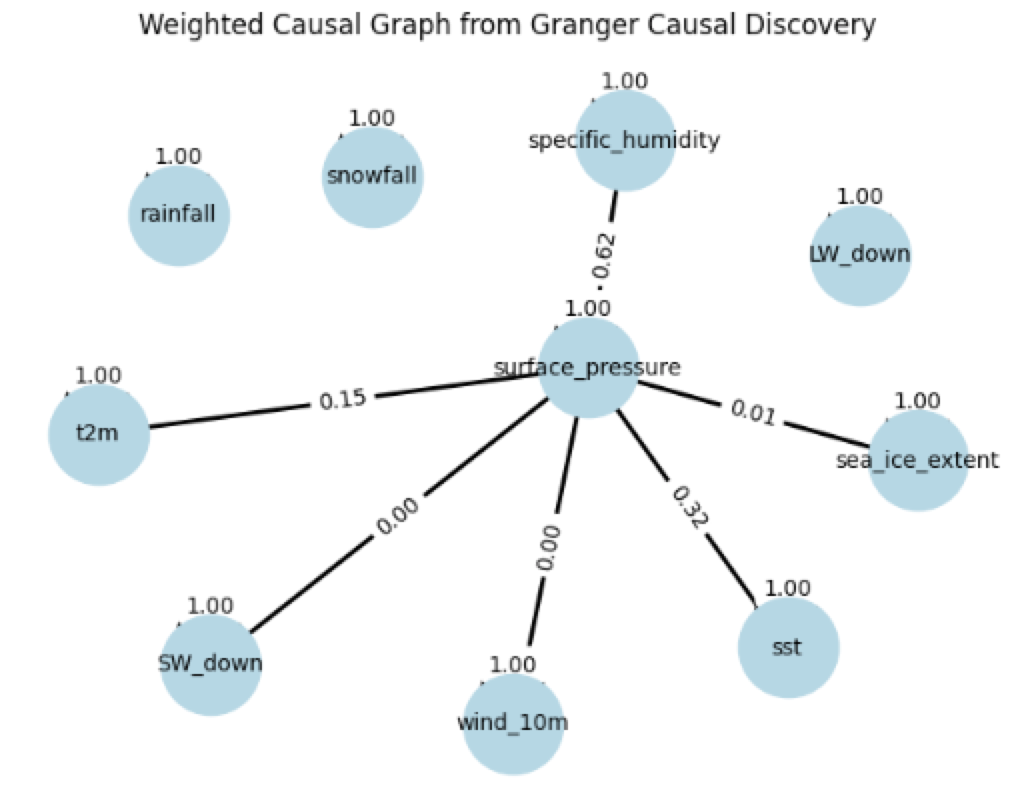}
   \caption{Granger causality results on Arctic dataset.}
   \Description{}
   \label{fig:rw_gc_graph}
\end{figure}
Figure~\ref{fig:rw_gc_graph} shows Granger detected only six significant edges, mostly linear lagged effects. This limited scope highlights its inability to capture nonlinear dependencies.

\subsubsection{PCMCI}
\begin{figure}[ht!]
   \centering
   \includegraphics[width=\linewidth]{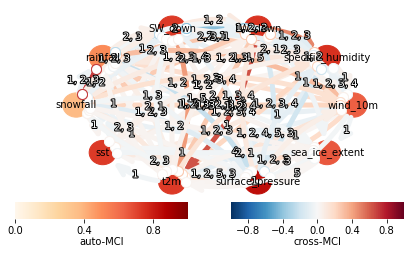}
   \caption{PCMCI causal graph on Arctic dataset.}
   \Description{}
   \label{fig:pcmci_r}
\end{figure}

PCMCI recovered a richer causal structure (Figure~\ref{fig:pcmci_r}), including multiple atmospheric interactions. However, several detected edges lacked physical interpretability, suggesting potential false positives.

\subsubsection{VarLiNGAM}
\begin{figure}[ht!]
   \centering
   \includegraphics[width=\linewidth]{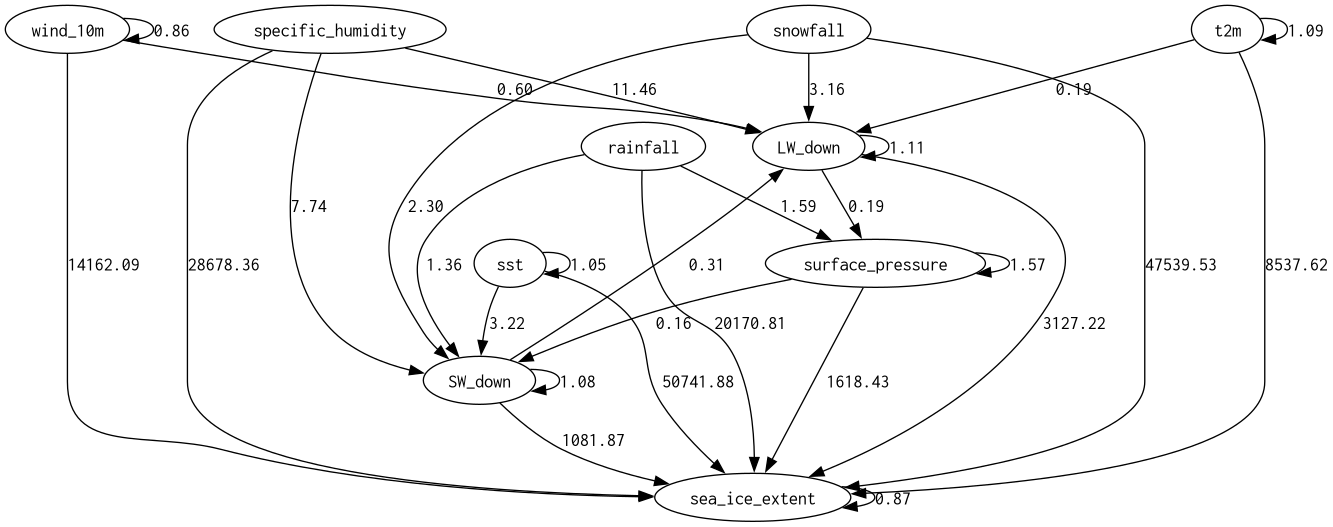}
   \caption{Learned causal graph from Arctic data using VarLiNGAM.}
   \Description{}
   \label{fig:VarLiNGAM_Real}
\end{figure}

VarLiNGAM identified robust linear causal relations including links between sea ice extent, wind, and humidity. While statistically strong, its scope remained restricted to linear effects.
Figure~\ref{fig:VarLiNGAM_Real} is the learned causal graph from real-world data. We considered a lower threshold of \(0.7\) meaning we considered edges with a causal effect larger than minimum threshold.

\subsubsection{CCM} 
\begin{figure}[htp!]
   \centering
   \includegraphics[width=\columnwidth]{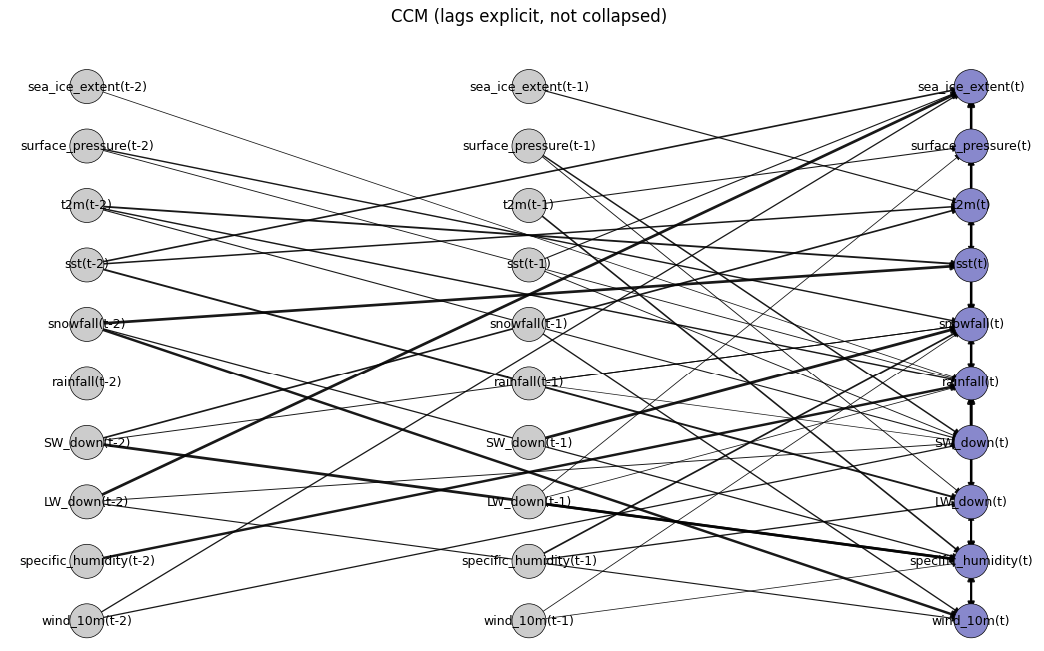}
   \caption{CCM learned causal graph on Arctic dataset.}
   \Description{}
   \label{fig:ccm-real}
\end{figure}

\begin{figure}[htp!]
   \centering
   \includegraphics[width=\columnwidth]{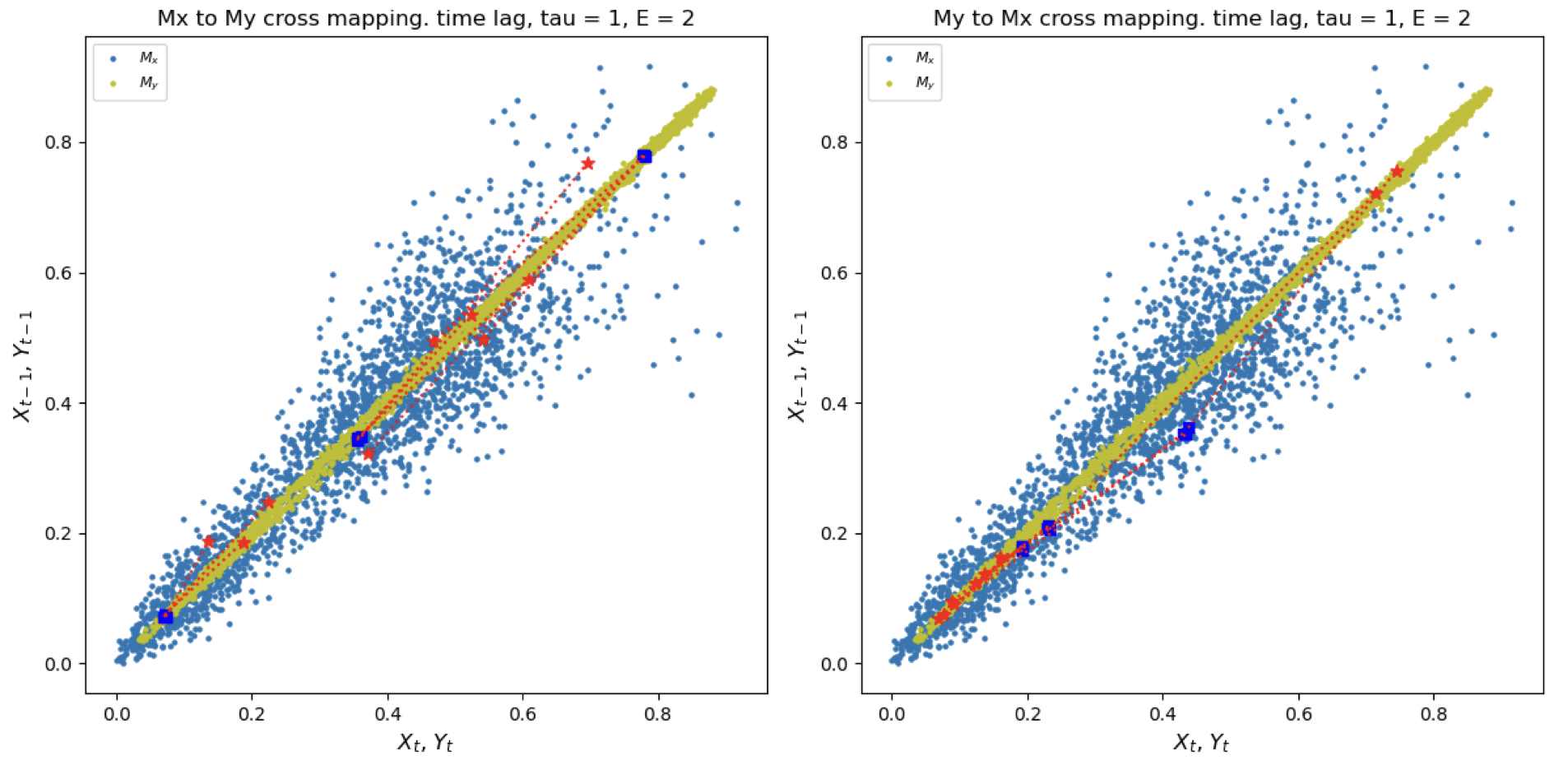}
   \caption{Cross mapping: snowfall $\sim$ sea ice extent.}
   \Description{}
   \label{fig:sihu1}
\end{figure}

\begin{figure}[htp!]
   \centering
   \includegraphics[width=\columnwidth]{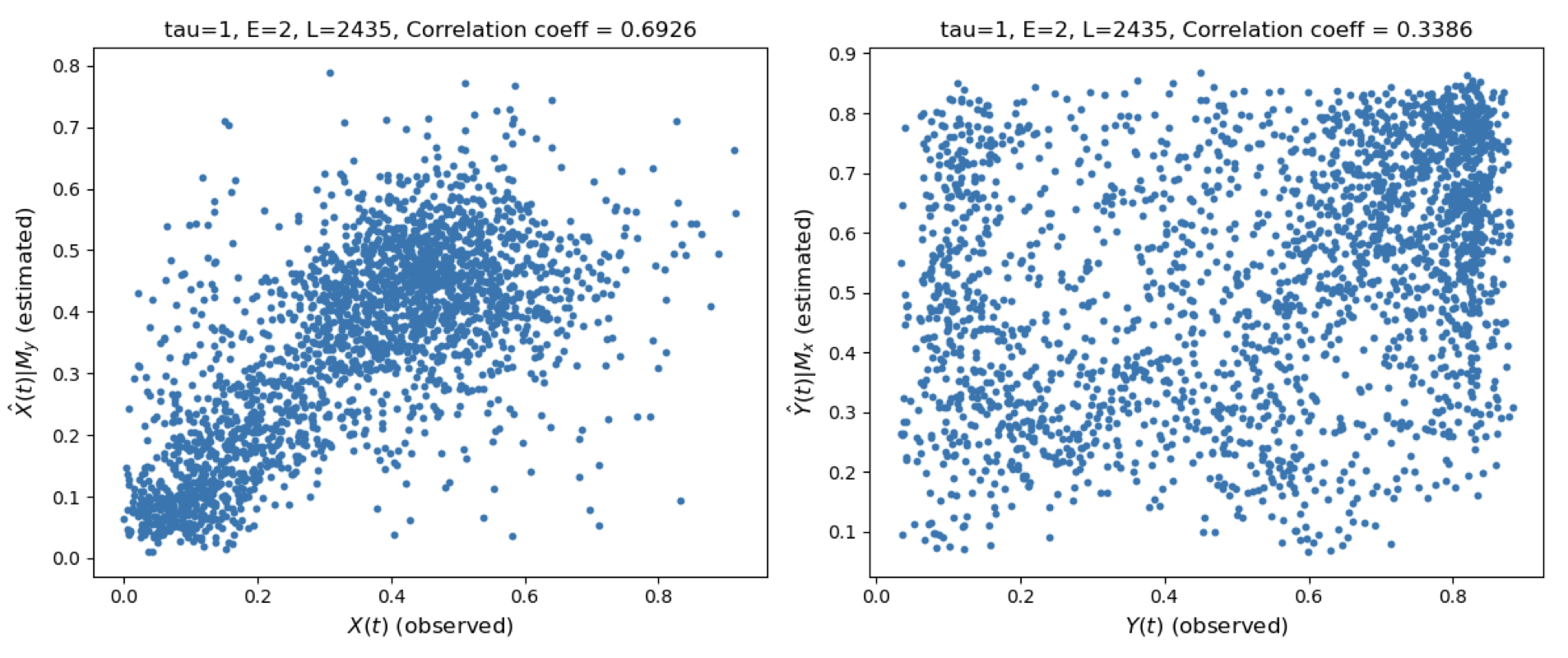}
   \caption{Visualize correlation between snowfall $\sim$ sea ice extent.}
   \Description{}
   \label{fig:sihu2}
\end{figure}

\begin{figure}[htp!]
   \centering
   \includegraphics[width=\columnwidth]{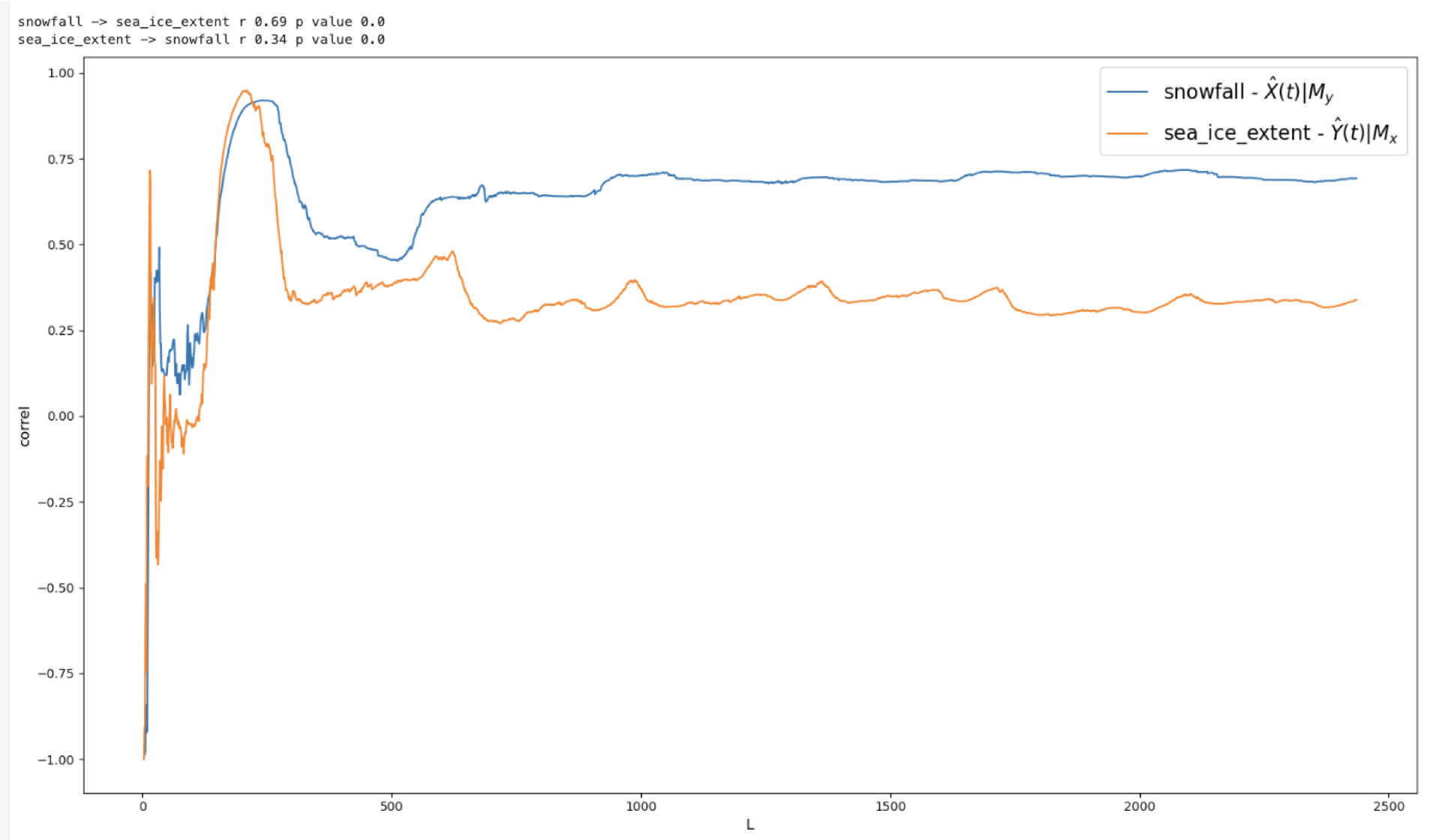}
   \caption{Convergence: snowfall $\sim$ sea ice extent.}
   \Description{}
   \label{fig:sihu3}
\end{figure}

Figure~\ref{fig:ccm-real} revealed robust and physically consistent feedbacks present in the Arctic sea ice data that remain undetected by conventional linear approaches.
 In particular, CCM identified bidirectional couplings between sea ice extent and key climate variables, including snowfall, humidity, longwave radiation, and near-surface temperature. The detection of the snowfall–sea ice extent feedback, illustrated in Figures~\ref{fig:sihu1}--\ref{fig:sihu3}, highlights the ability of CCM to uncover nonlinear interactions of critical relevance to Arctic climate dynamics.
There are two requirements for causal inference, narrow cross convergence and cross mapping strength of the observed variables often measured through correlation. Stronger causal influence between variables is reflected in narrower cross mappings. Figure~\ref{fig:sihu1} illustrates the cross mapping of points from one manifold (blue box) to the corresponding points on the other manifold (red star). We see that values from one manifold $\mathcal{M}_x$ generate narrowed cross mappings to values on another manifold $\mathcal{M}_y$, indicating that we can precisely predict $\mathcal{M}_y$ or $Y$ given $\mathcal{M}_x$. This suggests that much information from $Y$ is transferred to $X$, implying that $Y$ strongly drives $X$. 
Physically, heavy snow on sea ice acts as an insulating layer that reduces heat loss from the ocean to the atmosphere, thereby slowing ice growth and reducing extent (a negative forcing on ice growth). Conversely, changes in sea ice extent can modify surface albedo, temperature, and moisture transport, thus influencing snowfall patterns (a feedback effect) \cite{clemens2023snow}. This is further seen in Figure~\ref{fig:sihu2} when CCM is applied. In the left panel the performance of cross mapping from $M_y$ to $M_x$ for predicting $X(t)$ highlights that $X$ can be reliably inferred from $Y$. This indicates that $Y$ encapsulates sufficient dynamical information about $X$. Consequently, we infer that $X$ exerts a causal influence on $Y$. The key intuition is that if $X$ drives $Y$, then the historical dynamics of $Y$ inevitably encode traces of $X$, enabling robust causal inference through state-space reconstruction. For completeness, we report cross mapping in both directions; however, one direction consistently exhibits stronger predictive power, providing evidence of asymmetric causal dependence.

Figure~\ref{fig:sihu3} indicates that cross mapping accuracy improves as the time horizon $L$ at the x-axis increases.  
This behavior arises only when the reconstruction of the shared attractor between snowfall and sea ice extent becomes more accurate with additional data. We also observe that both cross mappings exhibit convergence, though to differing extents.  
Thus, the two fundamental criteria of Convergent Cross Mapping \textit{convergence} and \textit{reciprocal cross mapping} are satisfied, providing robust evidence of causal interaction.  
Importantly, the asymmetry in convergence magnitudes in both variables reveals that the strength of causality is direction-dependent, with snowfall influence exerting a stronger effect than sea ice extent. This highlights not only the existence of causality but also its relative intensity across directions, underscoring the discriminative power of CCM in capturing nuanced causal dynamics.  

\subsection{Comparative Performance Analysis}
Tables~\ref{tab:sieccm} and~\ref{tab:synccm} compare CCM and Granger using $p$-values. For the sea ice dataset (Table~\ref{tab:sieccm}), CCM detected significant feedbacks ($p=0.009$) for \textit{humidity}, \textit{LW\_down}, \textit{SST}, and \textit{t2m}, consistent with known Arctic amplification processes. Granger, in contrast, reported universal significance ($p=0.000$), indicating false positives.  

For the synthetic dataset (Table~\ref{tab:synccm}), CCM correctly identified $S_2 \leftrightarrow S_8$ as significant ($p=0.001$), while Granger’s results varied widely by lag order, again suggesting instability for nonlinear systems.

\begin{table}[ht!]
\centering
\caption{Comparison of $p$-values for \texttt{sea\_ice} dataset.}
\label{tab:sieccm}
\resizebox{\columnwidth}{!}{%
\begin{tabular}{|c|c|c|c|c|c|c|c|c|c|}
\hline
 & snow & rain & wind & hmdt & LW & SW & sst & t2m & Prsr \\ \hline
$\rho_{(\text{CCM})}$  & 0.999 & 0.999 & 0.198 & 0.009 & 0.009 & 0.999 & 0.009 & 0.009 & 0.784 \\ \hline
$\rho_{(\text{Grng})}$ & 0.000 & 0.000 & 0.000 & 0.000 & 0.000 & 0.000 & 0.000 & 0.000 & 0.000 \\ \hline
\end{tabular}%
}
\end{table}

\begin{table}[ht!]
    \centering
    \caption{Comparison of $p$-values for synthetic dataset.}
    \label{tab:synccm}
    \resizebox{\columnwidth}{!}{%
    \begin{tabular}{|c|c|c|c|c|c|c|c|}
        \hline
        & S1$\sim$S8 & S2$\sim$S8 & S3$\sim$S8 & S4$\sim$S8 & S5$\sim$S8 & S6$\sim$S8 & S7$\sim$S8 \\ \hline
        $\rho_{(\text{CCM})}$ & 0.491 & 0.001 & 0.874 & 0.322 & 0.645 & 0.871 & 0.155 \\ \cline{2-8}
        $\rho_{(\text{Grng})^{12}}$ & 0.000 & 0.000 & 0.001 & 0.012 & 0.000 & 0.000 & 0.281 \\ \hline
        $\rho_{(\text{Grng})^{2}}$ & 0.647 & 0.258 & 0.273 & 0.603 & 0.000 & 0.000 & 0.003 \\ \hline
    \end{tabular}%
    }
\end{table}

\noindent
\textbf{Summary.}  
Linear approaches effectively captured dominant direct dependencies but systematically failed to recover nonlinear feedback mechanisms. While PCMCI demonstrated superior scalability to high-dimensional settings, it frequently introduced spurious connections that undermined interpretability. In contrast, CCM consistently detected weak, bidirectional, and physically grounded causal relationships, highlighting its robustness and suitability for analyzing complex, coupled climate systems.

\section{Discussion}

The comparative evaluation of causal discovery methods on synthetic and Arctic climate data highlights clear differences in capability and suitability for climate research. Granger causality showed high sensitivity but poor specificity, particularly in real-world data where it assigned significance to nearly all variable pairs ($p=0.000$), reflecting overfitting and inability to distinguish genuine from spurious relationships. PCMCI improved robustness by incorporating conditional independence tests and controlling for confounders, recovering several plausible links. However, its reliance on linear assumptions limited its performance on nonlinear feedbacks. VarLiNGAM provided reliable detection of linear and lagged relationships with good specificity and physical interpretability (e.g., sea surface temperature driving near-surface air temperature). 

CCM exhibited unique strengths: it consistently identified weak nonlinear causal links and bidirectional feedback loops while avoiding false positives. On synthetic data, CCM detected only the designed causal connections. On Arctic data, it revealed meaningful relationships between sea ice extent and atmospheric variables (specific humidity, longwave radiation, and temperature), aligning with established ice–albedo and moisture–radiation feedback mechanisms central to Arctic amplification. These results confirm that method selection should depend on system characteristics: VarLiNGAM for primarily linear systems, PCMCI when confounding control is critical, and CCM for nonlinear, coupled systems with feedback dynamics.

While the findings validate theoretical understanding of Arctic climate processes, limitations remain. The synthetic dataset cannot fully represent the multiscale complexity of real climate systems, and this study focused only on the Arctic region. 
Future work should extend to diverse climatic contexts, explore hybrid approaches combining method strengths, and incorporate physical priors and uncertainty quantification to improve reliability of causal discovery in Earth system science.

\begin{table}[ht!]
    \centering
    \caption{Comparison of weak and strong causal links in Convergence Cross Mapping (CCM).}
    \label{tab:weakstrongccm}
    \resizebox{\columnwidth}{!}{%
    \begin{tabular}{|l|p{5cm}|p{5cm}|}
        \hline
        \textbf{Property} & \textbf{Weak Causal Link} & \textbf{Strong Causal Link} \\ \hline
        Cross map correlation ($\rho$) & Low (e.g., 0.2--0.5) & High (e.g., 0.7--1.0) \\ \hline
        Convergence behavior & Slow or noisy convergence of $\rho$ with library size $L$ & Clear monotonic convergence with $L$ \\ \hline
        Signal embedding & Weak imprint of the driver variable in the target manifold & Strong embedding of the driver’s dynamics in the target manifold \\ \hline
        System coupling & Loose or nonlinear coupling; influence modulated by other variables & Tight deterministic coupling between variables \\ \hline
        Causal interpretation & Potential influence or indirect causal pathway & Direct, dominant causal driver \\ \hline
    \end{tabular}%
    }
\end{table}

\section{Conclusions}
This study benchmarks causal discovery methods on synthetic and Arctic climate datasets. It compares Granger causality, PCMCI, VarLiNGAM, and CCM. Performance varies strongly with underlying system dynamics. Granger excels when dependencies are linear and strong. VarLiNGAM is also effective for linear, non-Gaussian structures. PCMCI improves robustness via conditional independence testing. CCM stands out in complex, nonlinear, and coupled systems. It can capture bidirectional feedbacks common in climate processes. Applied to Arctic observations, CCM revealed causal links involving sea ice extent. Specific humidity emerged as a key atmospheric driver. Longwave radiation showed strong causal interactions with sea ice. Near-surface temperature also exhibited influential feedback pathways. These findings provide quantitative evidence for mechanisms behind Arctic amplification. They highlight the necessity of nonlinear methods in climate causal discovery. Future work will extend CCM to high-dimensional settings, integrate hybrid linear–nonlinear frameworks, and apply them to other climate regions and processes.

\bibliographystyle{ACM-Reference-Format}
\bibliography{references}

@article{van2015causal,
  title={Causal feedbacks in climate change},
  author={Van Nes, Egbert H and Scheffer, Marten and Brovkin, Victor and Lenton, Timothy M and Ye, Hao and Deyle, Ethan and Sugihara, George},
  journal={Nature Climate Change},
  volume={5},
  number={5},
  pages={445--448},
  year={2015},
  publisher={Nature Publishing Group}
}

@article{stroeve2012arctic,
  title={The Arctic’s rapidly shrinking sea ice cover: a research synthesis},
  author={Stroeve, Julienne C and Serreze, Mark C and Holland, Marika M and Kay, Jennifer E and Malanik, James and Barrett, Andrew P},
  journal={Climatic change},
  volume={110},
  number={3},
  pages={1005--1027},
  year={2012},
  publisher={Springer}
}

@inproceedings{takens2006detecting,
  title={Detecting strange attractors in turbulence},
  author={Takens, Floris},
  booktitle={Dynamical Systems and Turbulence, Warwick 1980: proceedings of a symposium held at the University of Warwick 1979/80},
  pages={366--381},
  year={2006},
  organization={Springer}
}

@article{shimizu2006linear,
  title={A linear non-Gaussian acyclic model for causal discovery.},
  author={Shimizu, Shohei and Hoyer, Patrik O and Hyv{\"a}rinen, Aapo and Kerminen, Antti and Jordan, Michael},
  journal={Journal of Machine Learning Research},
  volume={7},
  number={10},
  year={2006}
}

@misc{nji2024evaluation,
  author = {F. N. Nji and O. Faruque and M. Cham and J. Vandana and J. Wang},
  title = {Evaluation of Traditional and Deep Clustering Algorithms for Multivariate Spatiotemporal Data},
  year = {2024},
  url = {https://www.osti.gov/servlets/purl/2519314}
}

@article{faruque2023deep,
  author = {Faruque, Omar and Nji, Francis Ndikum and Cham, Mostafa and Salvi, Rohan Mandar and Zheng, Xue and Wang, Jianwu},
  title = {Deep spatiotemporal clustering: A temporal clustering approach for multi-dimensional climate data},
  year = {2023},
  journal = {arXiv},
  volume = {},
  number = {},
  pages = {},
  doi = {10.48550/arXiv.2304.14541},
  url = {https://doi.org/10.48550/arXiv.2304.14541}
}

@INPROCEEDINGS{10825871,
  author={Nji, Francis Ndikum and Faruque, Omar and Cham, Mostafa and Vandana, Janeja and Wang, Jianwu},
  booktitle={2024 IEEE International Conference on Big Data (BigData)}, 
  title={Hybrid Ensemble Deep Graph Temporal Clustering for Spatiotemporal Data}, 
  year={2024},
  volume={},
  number={},
  pages={4374-4383},
  doi={10.1109/BigData62323.2024.10825871}}

@article{shojaie2022granger,
  title={Granger causality: A review and recent advances},
  author={Shojaie, Ali and Fox, Emily B},
  journal={Annual Review of Statistics and Its Application},
  volume={9},
  number={1},
  pages={289--319},
  year={2022},
  publisher={Annual Reviews}
}

@article{jiao2024optimizing,
  title={Optimizing VarLiNGAM for Scalable and Efficient Time Series Causal Discovery},
  author={Jiao, Ziyang and Guo, Ce and Luk, Wayne},
  journal={arXiv preprint arXiv:2409.05500},
  year={2024}
}

@article{runge2019inferring,
  title={Inferring causation from time series in Earth system sciences},
  author={Runge, Jakob and Bathiany, Sebastian and Bollt, Erik and Camps-Valls, Gustau and Coumou, Dim and Deyle, Ethan and Glymour, Clark and Kretschmer, Marlene and Mahecha, Miguel D and Mu{\~n}oz-Mar{\'\i}, Jordi and others},
  journal={Nature communications},
  volume={10},
  number={1},
  pages={2553},
  year={2019},
  publisher={Nature Publishing Group UK London}
}

@article{hoyer2008nonlinear,
  title={Nonlinear causal discovery with additive noise models},
  author={Hoyer, Patrik and Janzing, Dominik and Mooij, Joris M and Peters, Jonas and Sch{\"o}lkopf, Bernhard},
  journal={Advances in neural information processing systems},
  volume={21},
  year={2008}
  }

@article{barnett2009granger,
  title={Granger causality and transfer entropy are equivalent for Gaussian variables},
  author={Barnett, Lionel and Barrett, Adam B and Seth, Anil K},
  journal={Physical review letters},
  volume={103},
  number={23},
  pages={238701},
  year={2009},
  publisher={APS}
}

@online{CCM,
 author = {Francis},
 editor = {{Github}},
 
 url = {https://github.com/Frankie0609/Causal_AI/blob/main/CCM_Team_Project.ipynb},
 urldate = {10.29.2022},
 originalyear = {10.29.2022}
}

@article{ye2015distinguishing,
  title={Distinguishing time-delayed causal interactions using convergent cross mapping},
  author={Ye, Hao and Deyle, Ethan R and Gilarranz, Luis J and Sugihara, George},
  journal={Scientific reports},
  volume={5},
  number={1},
  pages={1--9},
  year={2015},
  publisher={Nature Publishing Group}
}

@article{clark2015spatial,
  title={Spatial convergent cross mapping to detect causal relationships from short time series},
  author={Clark, Adam Thomas and Ye, Hao and Isbell, Forest and Deyle, Ethan R and Cowles, Jane and Tilman, G David and Sugihara, George},
  journal={Ecology},
  volume={96},
  number={5},
  pages={1174--1181},
  year={2015},
  publisher={Wiley Online Library}
}

@article{wang2018detecting,
  title={Detecting the causal effect of soil moisture on precipitation using convergent cross mapping},
  author={Wang, Yunqian and Yang, Jing and Chen, Yaning and De Maeyer, Philippe and Li, Zhi and Duan, Weili},
  journal={Scientific reports},
  volume={8},
  number={1},
  pages={1--8},
  year={2018},
  publisher={Nature Publishing Group}
}

@article{vihma2014effects,
  title={Effects of Arctic sea ice decline on weather and climate: A review},
  author={Vihma, Timo},
  journal={Surveys in Geophysics},
  volume={35},
  number={5},
  pages={1175--1214},
  year={2014},
  publisher={Springer}
}

@article{stammerjohn2012regions,
  title={Regions of rapid sea ice change: An inter-hemispheric seasonal comparison},
  author={Stammerjohn, Sharon and Massom, Robert and Rind, David and Martinson, Douglas},
  journal={Geophysical Research Letters},
  volume={39},
  number={6},
  year={2012},
  publisher={Wiley Online Library}
}

@article{polyakov2012recent,
  title={Recent changes of Arctic multiyear sea ice coverage and the likely causes},
  author={Polyakov, Igor V and Walsh, John E and Kwok, Ronald},
  journal={Bulletin of the American Meteorological Society},
  volume={93},
  number={2},
  pages={145--151},
  year={2012},
  publisher={JSTOR}
}

@inproceedings{de2020latent,
  title={Latent convergent cross mapping},
  author={De Brouwer, Edward and Arany, Adam and Simm, Jaak and Moreau, Yves},
  booktitle={International Conference on Learning Representations},
  year={2020}
}

@inproceedings{runge2020discovering,
  title={Discovering contemporaneous and lagged causal relations in autocorrelated nonlinear time series datasets},
  author={Runge, Jakob},
  booktitle={Conference on Uncertainty in Artificial Intelligence},
  pages={1388--1397},
  year={2020},
  organization={Pmlr}
}

@article{dickey1979distribution,
  title={Distribution of the estimators for autoregressive time series with a unit root},
  author={Dickey, David A and Fuller, Wayne A},
  journal={Journal of the American statistical association},
  volume={74},
  number={366a},
  pages={427--431},
  year={1979},
  publisher={Taylor \& Francis}
}

@article{kwiatkowski1992testing,
  title={Testing the null hypothesis of stationarity against the alternative of a unit root: How sure are we that economic time series have a unit root?},
  author={Kwiatkowski, Denis and Phillips, Peter CB and Schmidt, Peter and Shin, Yongcheol},
  journal={Journal of econometrics},
  volume={54},
  number={1-3},
  pages={159--178},
  year={1992},
  publisher={Elsevier}
}

@inproceedings{ali2022benchmarking,
  title={Benchmarking probabilistic machine learning models for arctic sea ice forecasting},
  author={Ali, Sahara and Mostafa, Seraj AM and Li, Xingyan and Khanjani, Sara and Wang, Jianwu and Foulds, James and Janeja, Vandana},
  booktitle={IGARSS 2022-2022 IEEE International Geoscience and Remote Sensing Symposium},
  pages={4654--4657},
  year={2022},
  organization={IEEE},
  publisher={arXiv}
}

@inproceedings{su2023uncovering,
  title={Uncovering Causal Relationships in Co-location Patterns: Approximating Direct Causes through Granger Causality Mining},
  author={Su, Hong and Wong, Raymond Chi-Wing},
  booktitle={Proceedings of the 31st ACM International Conference on Advances in Geographic Information Systems},
  pages={1--2},
  year={2023}
}

@article{gao2023causal,
  title={Causal inference from cross-sectional earth system data with geographical convergent cross mapping},
  author={Gao, Bingbo and Yang, Jianyu and Chen, Ziyue and Sugihara, George and Li, Manchun and Stein, Alfred and Kwan, Mei-Po and Wang, Jinfeng},
  journal={nature communications},
  volume={14},
  number={1},
  pages={5875},
  year={2023},
  publisher={Nature Publishing Group UK London}
}

@article{blackport2019minimal,
  title={Minimal influence of reduced Arctic sea ice on coincident cold winters in mid-latitudes},
  author={Blackport, Russell and Screen, James A and van der Wiel, Karin and Bintanja, Richard},
  journal={Nature climate change},
  volume={9},
  number={9},
  pages={697--704},
  year={2019},
  publisher={Nature Publishing Group UK London}
}

@article{clemens2023snow,
  title={Snow loss into leads in Arctic sea ice: Minimal in typical wintertime conditions, but high during a warm and windy snowfall event},
  author={Clemens-Sewall, David and Polashenski, Chris and Frey, Markus M and Cox, Christopher J and Granskog, Mats A and Macfarlane, Amy R and Fons, Steven W and Schmale, Julia and Hutchings, Jennifer K and von Albedyll, Luisa and others},
  journal={Geophysical Research Letters},
  volume={50},
  number={12},
  pages={e2023GL102816},
  year={2023},
  publisher={Wiley Online Library}
}

@article{rial2004nonlinearities,
  title={Nonlinearities, feedbacks and critical thresholds within the Earth's climate system},
  author={Rial, Jos{\'e} A and Pielke Sr, Roger A and Beniston, Martin and Claussen, Martin and Canadell, JOsep and Cox, Peter and Held, Hermann and de Noblet-Ducoudr{\'e}, Nathalie and Prinn, Ronald and Reynolds, James F and others},
  journal={Climatic change},
  volume={65},
  number={1},
  pages={11--38},
  year={2004},
  publisher={Springer}
}

\end{document}